\documentclass{aastex62}

\received{July 19, 2018}
\revised{March 20, 2019}
\accepted{April 4, 2019}
\submitjournal{ApJ}

\shorttitle{X-ray properties of Far-IR selected DOGs at z$\sim$2-3 in the COSMOS field}
\shortauthors{Riguccini et al.}

\begin{document}

\title{The composite nature of Dust-Obscured Galaxies (DOGs) at z$\sim$2-3 in the COSMOS field: II. The AGN fraction}

\correspondingauthor{Laurie A. Riguccini}
\email{riguccini@astro.ufrj.br}

\author[0000-0001-8343-6835]{Laurie A. Riguccini}
\affil{Observat\'orio do Valongo, Universidade Federal do Rio de Janeiro, Ladeira do Pedro Ant\^onio 43, Sa\'ude, \\ Rio de Janeiro, RJ 20080-090, Brazil }
\affiliation{CAPES/BJT Science Without Borders Postdoctoral Fellow, Brazil}

\author{Ezequiel Treister}
\affiliation{Instituto de Astrof\'{\i}sica, Facultad de F\'{i}sica, Pontificia Universidad Cat\'{o}lica de Chile, Casilla 306, Santiago 22, Chile}

\author{Karin Men\'endez-Delmestre}
\affiliation{Observat\'orio do Valongo, Universidade Federal do Rio de Janeiro, Ladeira do Pedro Ant\^onio 43, Sa\'ude, \\ Rio de Janeiro, RJ 20080-090, Brazil }

\author{Carolin Cardamone}
\affiliation{Math and Science Department, Wheelock College, Boston, MA 02215, USA}

\author{Francesca Civano}
\affiliation{Harvard-Smithsonian Center for Astrophysics, Cambridge, MA 02138, USA}
 
\author{Thiago S. Gon\c calves}
\affiliation{Observat\'orio do Valongo, Universidade Federal do Rio de Janeiro, Ladeira do Pedro Ant\^onio 43, Sa\'ude, \\ Rio de Janeiro, RJ 20080-090, Brazil }

\author{Guenther Hasinger}
\affiliation{Institute for Astronomy, University of Hawaii, 2680 Woodlawn Drive, Honolulu, HI 96822, USA}

\author{Anton M. Koekemoer}
\affiliation{Space Telescope Science Institute, 3700 San Martin Drive, Baltimore, MD 21218, USA}

\author{Giorgio Lanzuisi}
\affiliation{Dipartimento di Fisica e Astronomia, Universita di Bologna, viale Berti Pichat 6/2, 40127 Bologna, Italy}
\affiliation{INAF Osservatorio Astronomico di Bologna, Via Ranzani 1, 40127, Bologna, Italy}

\author{Emeric Le Floc'h}
\affiliation{Laboratoire AIM, CEA/DSM-CNRS-Universit\'e Paris Diderot, IRFU/Service d'Astrophysique, B\^at.709, CEA-Saclay, \\
91191 Gif-sur-Yvette Cedex, France \\}

\author{Elisabeta Lusso}
\affiliation{Centre for Extragalactic Astronomy, Department of Physics, Durham University, South Road, Durham, DH1 3LE, UK}

\author{Dieter Lutz}
\affiliation{Max-Planck-Institut f\"ur extraterrestrische Physik, Postfach 1312, Giessenbachstrasse 1, 85741 Garching, Germany}

\author{Stefano Marchesi} 
\affiliation{Department of Physics \& Astronomy, Clemson University, Clemson, SC 29634, USA}

\author{Takamitsu Miyaji}
\affiliation{Universidad Nacional Aut\'onoma de M\'exico sede Ensenada, Carret. Tijuna-Ensenada Km 103, Ensenada 22860, BC, Mexico}

\author{Francesca Pozzi}
\affiliation{INAF, Osservatorio Astronomico di Bologna, Via Ranzani 1, 40127 Bologna, Italy}
\affiliation{Dipartimento di Fisica e Astronomia, Universit\'a degli Studi di Bologna, Viale Berti Pichat 6/2, 40127 Bologna}

\author{Claudio Ricci}
\affiliation{N\'ucleo de Astronom\'ia de la Facultad de Ingenier\'ia, Universidad Diego Portales, Av. Ej\'ercito Libertador 441, Santiago, Chile}
\affiliation{Instituto de Astrof\'{\i}sica, Facultad de F\'{i}sica, Pontificia Universidad Cat\'{o}lica de Chile, Casilla 306, Santiago 22, Chile}
\affiliation{Kavli Institute for Astronomy and Astrophysics, Peking University, Beijing 100871, China }
\affiliation{Chinese Academy of Sciences South America Center for Astronomy and China-Chile Joint Center for Astronomy,\\ Camino El Observatorio 1515, Las Condes, Santiago, Chile}

\author{Giulia Rodighiero}
\affiliation{Dipartimento di Fisica e Astronomia ``G. Galilei'', Universita di Padova, Vicolo dell?Osservatorio 3, I-35122, Italy}

\author{Mara Salvato}
\affiliation{Max-Planck-Institute f\"ur Plasma Physics, Boltzmann Strasse 2, Garching 85748, Germany}

\author{Dave Sanders}
\affiliation{Institute for Astronomy, University of Hawaii, 2680 Woodlawn Drive, Honolulu, HI 96822, USA}

\author{Kevin Schawinski}
\affiliation{Institute for Astronomy, Department of Physics, ETH Zurich, Wolfgang-Pauli-Strasse 27, CH-8093 Zurich, Switzerland}

\author{Hyewon Suh}
\affiliation{Subaru Telescope, National Astronomical Observatory of Japan, 650 North A'ohoku Place, Hilo, HI, 96720, USA}



\begin{abstract}

We present the X-ray properties of 108 Dust-Obscured Galaxies (DOGs; F$_{24 \mu m}$/F$_{R} >$ 1000) in the COSMOS field, all of which detected in at least three far-infrared bands with the {\it Herschel} Observatory. Out of the entire sample, 22 are individually detected in the hard 2-8 keV X-ray band by the Chandra COSMOS Legacy survey,  
allowing us to classify them as AGN.
Of them, 6 (27\%) are Compton Thick AGN candidates with column densities N$_{H}$$>$10$^{24}$\,cm$^{-2}$ while 15 are moderately obscured AGNs with 10$^{22}$ $<$ N$_{H}$ $<$ 10$^{24}$\,cm$^{-2}$. Additionally, we estimate AGN contributions to the IR luminosity (8-1000$\mu$m rest-frame) greater than 20\% for 19 DOGs based on SED decomposition using {\it Spitzer}/MIPS 24$\mu$m and the five {\it Herschel} bands (100-500 $\mu$m). Only 7 of these 
are detected in X-rays individually. We performed a X-ray stacking 
analysis for the 86 undetected DOGs. We find that the AGN fraction in DOGs increases with 24$\mu$m flux and that it is higher than that of the 
general 24$\mu$m population. However, no significant difference is found when considering only X-ray detections. This strongly motivates the combined use of X-ray and far-IR surveys to successfully probe a wider population of AGNs, particularly for the most obscured ones.

\end{abstract}



%
%

\section{Introduction}
\label{sec:intro}

Pioneering work with the Infrared Astronomical Satellite (IRAS) and the Infrared Space Observatory (ISO) established that at low redshifts the most luminous infrared (IR) 
sources tend to be increasingly dominated by active galactic nuclei (AGNs; \citealp{Lutz:98,Genzel:98}). However, at higher redshifts the high luminosity of 
ultra-luminous IR galaxies (ULIRGs; L$_{IR}>$$\times$10$^{12}$\,L$\odot$) is not yet fully understood and significant diversity in the AGN-to-starburst 
ratio \citep[e.g.,][]{Sanders:99,Joseph:99,Desai:07,Karin:09,Sani:10,Petric:11,Pozzi:12} remains a critical difficulty in our understanding
of these sources. In an effort to address this and other questions, \citet{Dey:08} put forward color-based criteria to efficiently define a statistically-significant sample of dusty ULIRGs at z$\sim$1.5-3. By taking advantage of the unprecedented sensitivity and angular resolution at IR wavelengths of the {\it Spitzer Space Telescope}, they selected a population of optically-faint (22$<R<$27) and mid-IR bright (F$_{24\mu m}$ $>$  0.3 mJy) ``Dust Obscured Galaxies" (DOGs), defined as those sources having F$_{24 \mu m}$/F$_{R}$$>$1000.

The efficient selection of dust-obscured sources at high redshifts also had great impact on the search for hidden AGNs. Although X-ray surveys are a 
powerful tool to select unobscured and mildly-obscured AGNs, the current census of actively-growing supermassive black holes remains far from 
complete \citep[e.g,][]{Treister:04,Worsley:05,Tozzi:06,Page:06,Fiore:09,Juneau:11,Juneau:13}. The most obscured AGNs, in particular the deeply-embedded 
ones, are mostly absent in X-ray surveys. At these high column densities, the attenuation of X-rays is mainly due to Compton-scattering rather than 
photoelectric absorption; these sources are the so-called ``Compton-thick'' (CT) AGNs (N$_H$ $\simeq$ 1.5 $\times$ 10$^{24}$\,cm\,$^{-2}$), of which only a 
few have been identified in the local Universe \citep[][and references therein]{Burlon:11, Ricci:15}. At higher redshifts, hundreds of CT AGN 
candidates have been identified in X-rays thanks to XMM and Chandra observations at E$<$10 keV 
\citep[e.g.,][]{Comastri:11, Feruglio:11, Brightman:14,Buchner:15,Baronchelli:17} and NuSTAR data at higher X-ray energies \citep[e.g.,][]{Civano:15,Mullaney:15,Lansbury:17}. 

While in principle CT sources are just the high obscuration end of the AGN population, recent studies have shown that they might represent
a different and fundamental stage in setting up the super massive black hole (SMBH) growth-galaxy evolution connection. Indeed, \citet{Ricci:17} shows that there is a clear excess in the
relative number of CT AGN in the last stages of major galaxy mergers, consistent with this being one of the early phases of rapid SMBH growth
triggered by a major galaxy merger. Furthermore, a significant fraction of missed Compton thick accretion might hide an important part of the census of
SMBH growth across cosmic history. Indeed, a significant fraction of heavily obscured and CT sources are invoked at all redshifts in order to 
reproduce the observed Cosmic X-ray Background (CXRB) at 20-30 keV \citep[e.g.,][]{Comastri:95,Gilli:01,Ueda:03,Treister:05,Ballantyne:06,Gilli:07,Ueda:14}. 
However, the exact number of CT sources required by the CXRB is still heavily debated and ranges from $\sim$30\% \citep{Gilli:07} to 
$\sim$10\% \citep{Treister:09,Ballantyne:11}. Hence, determining the space density of CT AGN remains a critical open issue in our understanding
of the role of SMBH for galaxy evolution. Although a large proportion of the obscured AGN population still remains undetected, these objects can already account for a significant fraction of the total SMBH growth ($\sim$70\%; \citealt{Treister:05}). Indeed, AGN synthesis models that can 
explain the spectral shape and intensity of the CXRB predict a large volume density of heavily obscured and CT AGNs to reconcile the ``active'' and ``relic'' SMBH mass functions \citep[e.g.,][]{Gilli:01,Treister:04,Marconi:04,Treister:05,Gilli:07,Akylas:12}.

Since it is clear then that X-ray surveys are not sufficient to probe the complete AGN population, alternative selection 
techniques have been developed. Recent work by \citet{Riguccini:15} showed that a sub-sample of DOGs with far-IR (100--500$\mu$m) detection have a significant contribution from AGN activity at higher luminosities \citep{Riguccini:15}. This is consistent with recent work on mid-to-far IR Spectral Energy Distributions (SEDs) of luminous AGNs that have found that a higher AGN contribution in the far-IR, particularly at high AGN luminosities \citep[e.g.,][]{Symeonidis:16,Symeonidis:17}.
Because they are selected based on their far-IR output $-$ i.e., at longer wavelengths than the AGNs selected by near-through-mid IR surveys $-$ far-IR selected DOGs can potentially represent a distinctly-defined population of AGN candidates. 

Previous studies \citep[e.g.,][]{Fiore:09,Treister:09b} have focused on selecting sizable samples of high luminosity CT AGNs to measure accurately their 
volume density and to understand whether their obscuration properties are similar to those of lower luminosity AGNs.
In this work we adopt the following approach: based on a far-IR selection of DOGs with information on their AGN contribution 
(from a far-IR perspective) we exploit the Chandra COSMOS Legacy Survey \citep{Civano:16} to assess the AGN fraction in DOGs using the most recent and exquisite combination of far-IR and X-ray data. Our main aim is to quantify the AGN fraction in this population of far-IR DOGs using a multi-wavelength approach based on X-ray flux measurements and broad-band SED fitting.

The paper is organized as follows: we describe our data in Sect.  \ref{sec:dat} and our results in Section \ref{sec:results}. More detailed analysis and discussion
are presented in Section \ref{sec:discussion} while our conclusions are summarized in section \ref{sec:ccl}. Throughout this paper we assume a 
$\Lambda$CDM cosmology with H$_{0}$=70 km s$^{-1}$, $\Omega_m$ = 0.3, and $\Omega_{\Lambda}$ = 0.7. Unless otherwise specified, magnitudes 
are given in the AB system.

%
%

\section{Data}
\label{sec:dat}

\subsection{Far-IR}

As a reference to build our parent sample, we use the catalogues provided by the PEP and HerMES Herschel surveys \citep{Berta:11,Roseboom:10} to identify far IR-selected DOGs in the COSMOS field, detected in at least 3 of the 5 Herschel bands (cf Table\,\ref{tab:nb_sources} for a detailed explanation of the different selections). 
Those catalogues calculate source fluxes in each of the five Herschel bands by performing point spread function fitting at the positions of the 24$\mu$m-detected sources from \citet{Emeric:09}. Our total sample hence contains 108 far-IR detected DOGs. 

\begin{figure}
 \centering
 \resizebox{1.\hsize}{!}{\includegraphics{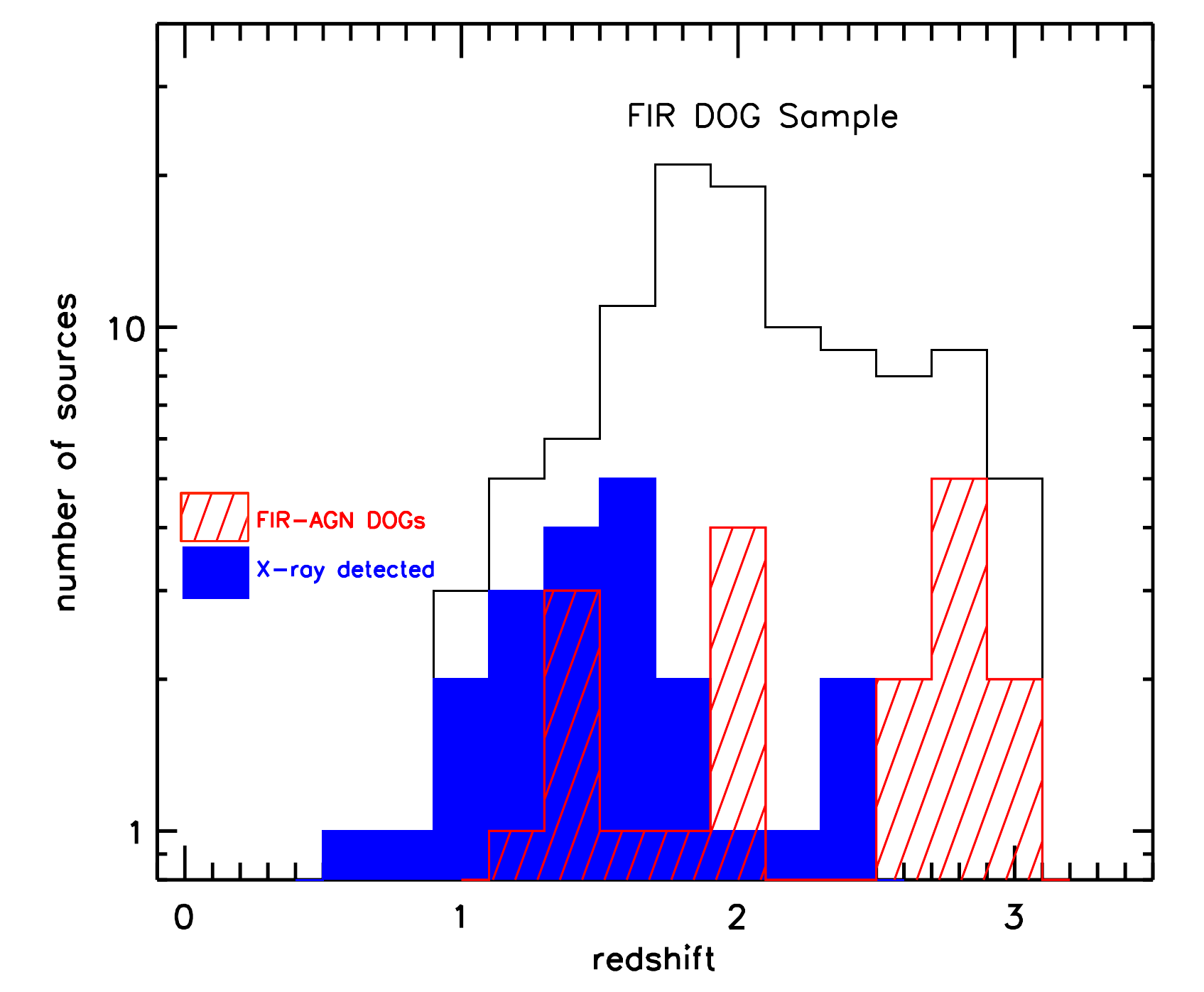}}
\caption{Redshift distribution (22 spectroscopic redshifts and 86 photometric redshifts) for the Herschel-selected DOG sample (\textit{black histogram}). 
The \textit{blue filled histogram} shows the sub-population of DOGs detected in X-rays with the COSMOS Chandra Legacy Survey, while the \textit{red hatched histogram} 
shows the sub-population of DOGs classified as AGN according to the mid-to-far-IR SED fitting. A KS test performed to these last two distributions indicates that they
are drawn from a different parent population with a probability of 6\,$\times$\,10$^{-3}$.}
 \label{fig:redshift_distrib}
\end{figure}

Among our sample of 108 far-IR selected DOGs, 22 sources have spectroscopic redshifts from Salvato et al. (in prep.). For the rest of our sample, we use the photometric redshifts determined by \citet{Riguccini:15} for their 95 DOGs based on SED fitting using the optical catalog of \citet{Ilbert:09} to access the photometry of these sources in the optical bands. We re-analyze the AGN contributions for these 22 DOGs using their spectroscopic redshifts, based on the approach described in \citet{Riguccini:15}. Fig.\,\ref{fig:redshift_distrib} shows the redshift distribution of our sample of 108 far-IR detected DOGs, which include 22 DOGs with spectroscopic redshifts and 86 DOGs with photometric redshifts.

\subsection{X-ray Data}

We use the Chandra COSMOS Legacy Survey \citep{Civano:16} to obtain the X-ray counterparts for the far-IR sources in our sample. The Chandra COSMOS 
Legacy Survey covers a total area of $\sim$2.2 deg$^2$, uniformly covering the $\sim$1.7 deg$^2$ COSMOS/HST field at  a $\sim$160\,ksec depth, expanding on the deep C-COSMOS area (1.45 vs 0.44 deg$^2$) by a factor 
of $\sim$3 at $\sim$3$\times$10$^{16}$ erg cm$^{-2}$ s$^{-1}$. The deeper and wider 
coverage of the Chandra COSMOS Legacy survey compared to previous X-ray observations of the COSMOS field \citep[e.g.,][]{Brusa:10,Brusa:07,Salvato:09}
allows us to detect new X-ray DOGs that have been missed by previous X-ray surveys. 

From a two arc-second cross-match between the 108 far-IR detected DOGs and the Chandra COSMOS Legacy data \citep{Civano:16} we identify X-ray counterparts for 22 of the sources in our sample, with a median X-ray flux of $\sim$10$^{-16}$ erg~s$^{-1}$ in the soft band (0.5-2 keV). From these 22, 9 are detected in X-rays for the first time thanks to the increased field coverage of the Chandra COSMOS Legacy Survey. \citet{Riguccini:15} associated Herschel sources with their optical counterparts using \citet{Ilbert:09} and had then access to the ID from Capak et al. (2007); see \citet{Riguccini:15} for details on the matching method.  As a sanity check, we cross-matched our results with the multi wavelength catalog (X-ray to near-IR) from \citet{Marchesi:16} and found the same optical ID.

\section{Results}
\label{sec:results}

\subsection{Source Classification}

Taking advantage of the far-IR data, the AGN and host galaxy contributions to the total IR flux have been constrained by \citet{Riguccini:15} for 95 out of the 108 
far-IR selected DOGs. They use the IDL-based SED-fitting procedure DecompIR, detailed in \citet{Mullaney:11} and combine 8 host-galaxy templates detailed in \citet{Riguccini:11} with an average AGN template. The validity of this procedure and of the AGN contributions to the IR luminosity obtained are discussed in \citet{Mullaney:11} and \citet{Riguccini:11}. \citet{Riguccini:15} found that 75\% of the far-IR DOGs are consistent with being dominated by star formation, while 16\% have a far-IR output with a significant contribution from an AGN (i.e. contribution from an AGN to the host galaxy$>$20\%). The SED fitting procedure failed for their remaining 9 DOGs (out of their sample of 95 DOGs), probably due to uncertainties in redshift, even after probing the different possibilities indicated by the PDF. We note that \citet{Riguccini:15} focused their work on the subsample of DOGs {\it a priori} associated with star formation, systematically excluding the $<4\%$ DOGs already-known to be AGNs with X-ray detections down to a flux limit of S$_{0.5-2 keV}$ = 5 $\times$ 10$^{-16}$ erg cm$^2$ s$^{-1}$.  For the remainder of the paper DOGs that are dominated by star-formation following the SED-fitting decomposition procedure (i.e. labelled ``host'' in Table\,\ref{tab:Xray_prop}) will be named SF-DOGs while the DOGs with a 20\% contribution to the IR 8-1000~$\mu$m luminosity, derived according to the SED-fitting decomposition procedure (i.e. labelled ``AGN'' in Table\,\ref{tab:Xray_prop}), will be referred as far-IR AGN-DOGs. 

We decompose the far-IR SED of the 22 X-ray detected DOGs into AGN and host galaxy components following the procedure described in \citet{Riguccini:15}. Amongst these 22 X-ray DOGs, 9 of them have been included in \citet{Riguccini:15}; however, taking advantage of the recent availability of spectroscopic redshifts for 22 of these we re-analyzed the SEDs of these sources. In the case of source DOG11 the re-analysis allowed for a satisfactory SED decomposition \citep[in contrast with][]{Riguccini:15}, enabling us to classify it as dominated by a host galaxy component.

We find that only 7 out of the 22 X-ray detected DOGs are classified as AGNs based on their far-IR SED, i.e. AGN fraction $>$20\%, cf \citet{Riguccini:15}. With the exception of one source that could not be properly decomposed using this procedure most likely due to a wrong redshift, the remaining X-ray detected DOGs (2/3 of the sample) are all classified as dominated by a host galaxy SED component.

 \begin{table}
\caption{Number of sources and characteristics of each selection described in this paper} 
\centering
    \begin{tabular}{ | l | p{7cm} |}
    \hline
    108 & DOGs with $F_{24 \mu m}>$80$\mu$Jy and with a 3-$\sigma$ detection in the 2 PACS bands and with a 3-$\sigma$ detection in at least one of the 3 SPIRE bands \\
 &  {\bf  Sample used for the remainder of the paper} \\
 & Hereafter far-IR DOGs
\\ \hline
   19 & far-IR AGN-DOGs following the far-IR SED-fitting analysis from \citet{Riguccini:15} \\ \hline
   83 & far-IR DOGs with $R-K>2.79$ \citep[e.g.,][]{Fiore:08,Fiore:09} \\ \hline
 2 & far-IR AGN-DOGs but with $R-K<2.79$ \\ \hline
 22 & far-IR DOGs with an X-ray detection in the Chandra COSMOS Legacy survey \citep{Civano:16} \\ \hline
 7 & far-IR DOGs with an X-ray detection in the Chandra COSMOS Legacy survey \citep{Civano:16} have been classified as AGN-DOGs following the same procedure than in \citet{Riguccini:15} \\ \hline
 6 & potential Compton Thick AGN (N$_H$ $>$ 10$^{24}$ cm$^{-2}$) but only 2 of them are labelled far-IR AGN DOGs \\ \hline
     \end{tabular}
\label{tab:nb_sources}
\end{table}

\begin{figure}
 \centering
 \resizebox{1.1\hsize}{!}{\includegraphics{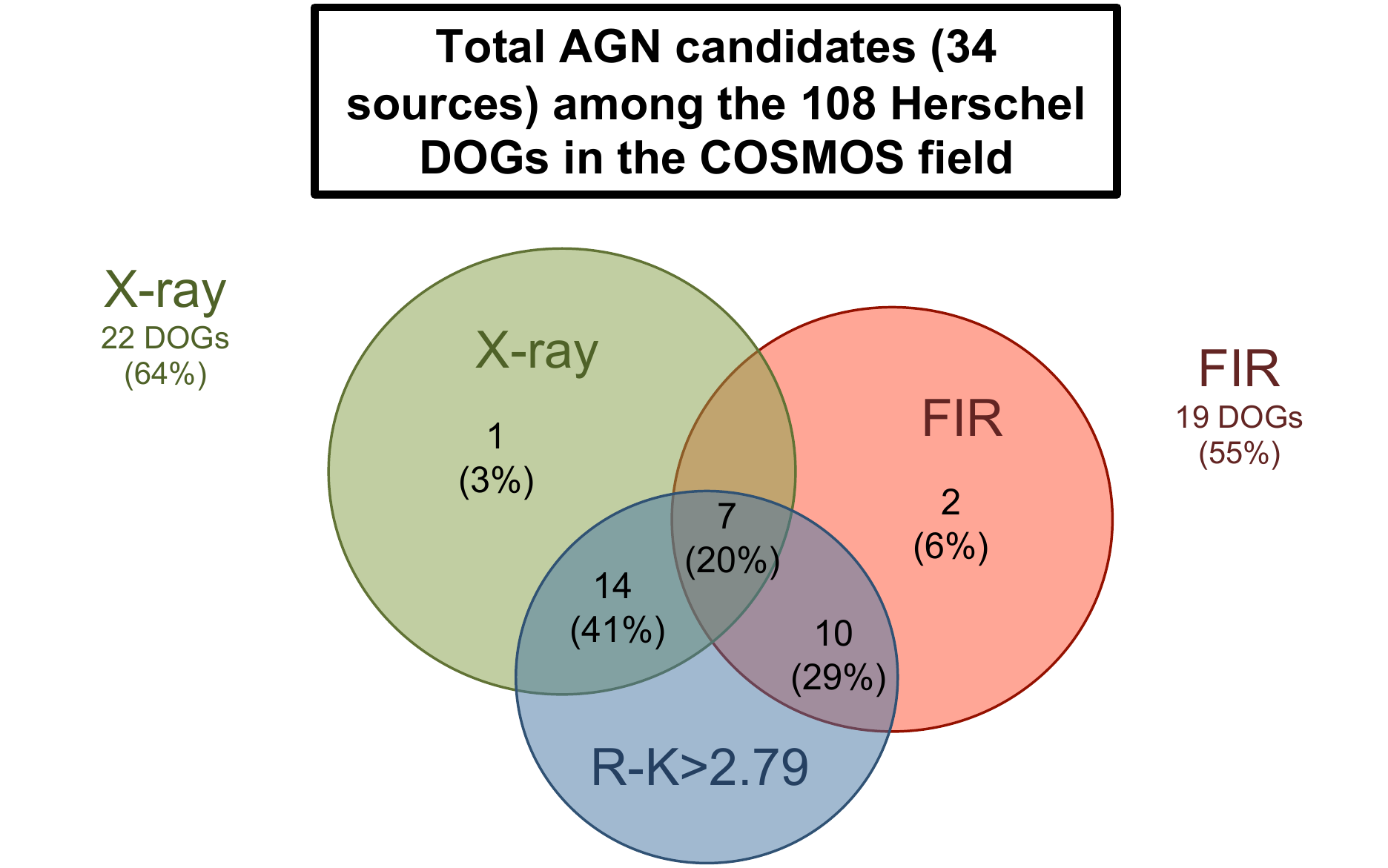}}
  \caption{Venn diagram showing the distribution of AGN selections within the DOG population detected with Herschel in the COSMOS field: X-ray selected AGN-DOGs in green, FIR selected AGN-DOGs (based on SED-fitting)  in red and R-K cut in blue. The numbers and percentages are given with respect to the whole AGN population considered in this work (i.e. X-ray + FIR).}
 \label{fig:Venn_diagram_DOGs}
\end{figure}

A summary of the AGN selections and the overlap between the different AGN criteria can be found on the Venn diagram presented on Fig.\,\ref{fig:Venn_diagram_DOGs}. The numbers are expressed with respect to the total number of AGN candidates among the Herschel DOG population, i.e. the AGN candidates selected from a hard X-ray detection and from the SED-fitting decomposition using FIR data.

\subsection{X-ray Properties}

\begin{figure}
 \centering
 \resizebox{1.\hsize}{!}{\includegraphics{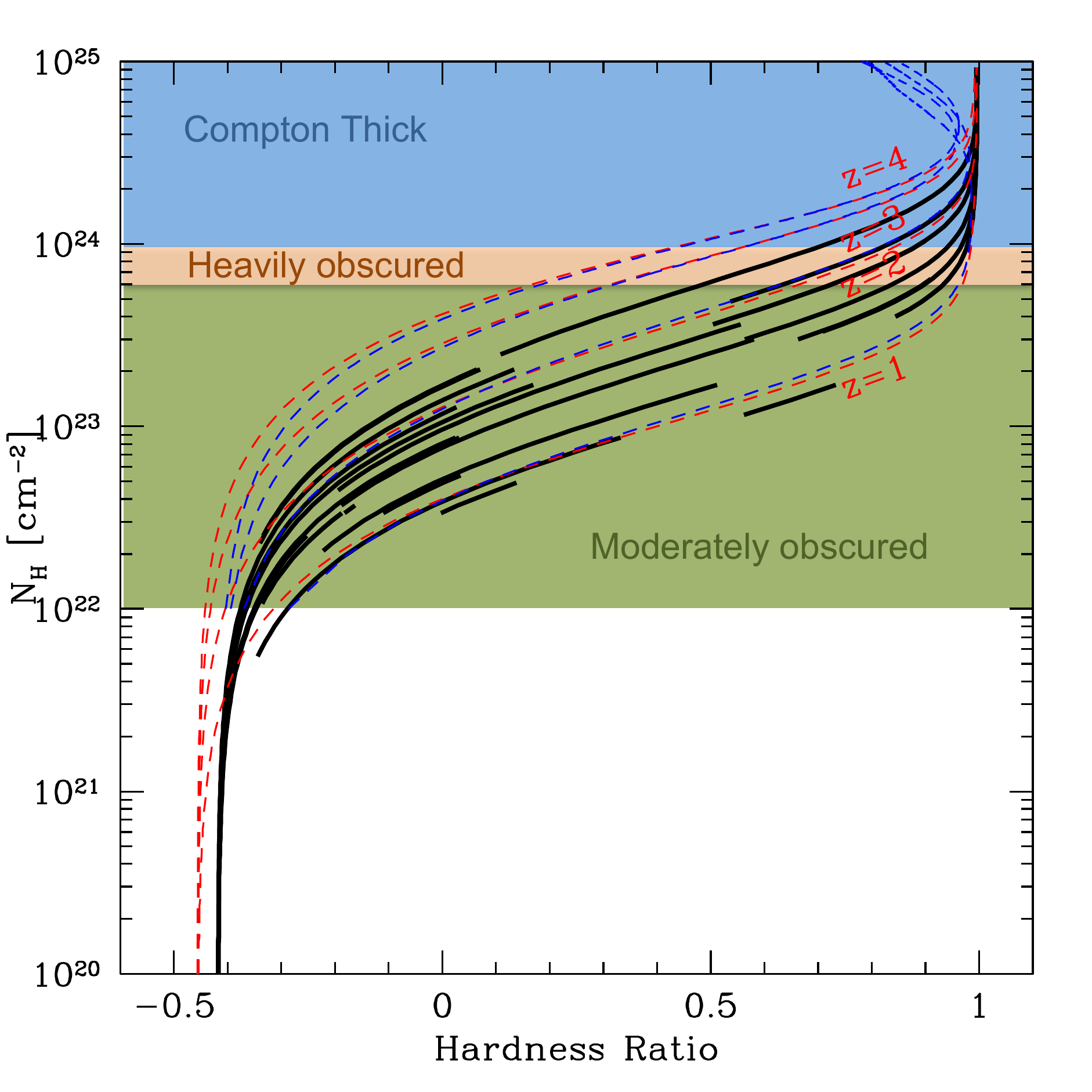}}
  \caption{Neutral hydrogen column density (N$_{H}$) as a function of hardness ratio (HR) for the X-ray detected DOGs. The blue region (with N$_{H}>$10$^{24}$ 
  cm$^{-2}$) represents the Compton-Thick (CT) population, while the orange one shows the location of heavily obscured sources with N$_{H}\sim$5$\times$10$^{23}$ 
  cm$^{-2}$ and the green region situated right below corresponds to the moderately obscured sources, with $N_H$(cm$^{-2}$)$>$10$^{22}$. The {\it dashed lines}
  show the expected relation between $N_H$ and HR for sources at $z$=1,2,3 and 4 assuming an intrinsic power law plus photoelectric absorption ({\it red lines}) and
  the MYTorus models ({\it blue lines}). The {\it black solid segments} show the location for the sources in Table\,\ref{tab:Xray_prop} considering the observed HR
  and their uncertainties and assuming the simple powerlaw model.}
 \label{fig:hr}
\end{figure}

The X-ray properties of the 22 DOGs individually detected by Chandra are described in Table\,\ref{tab:Xray_prop}. Given the faint X-ray fluxes,
which yield a relatively low number of counts, detailed fitting to the observed X-ray spectrum is not possible for the majority of these sources. However, we can estimate
the neutral hydrogen column density along the line of sight ($N_H$) from the observed X-ray count rate, following the procedure described by \citet{Treister:09b}. Briefly,
this is done by assuming that the intrinsic spectrum is a power-law with spectral index $\Gamma$=1.9 $-$ in agreement with the observed average AGN spectrum (e.g., \citealp{Nandra:94}) $-$ and computing the expected hardness ratio (HR). In the case of {\it Chandra}, the observed HR is defined as ($H$-$S$)/($H$+$S$), with  $S$ defined as the count rate in the soft X-ray band (0.5-2 keV) and $H$ as the count rate in the hard band (2-8 keV). The expected 
HR is computed for each source individually considering the redshift of the source and a range in photoelectric absorption parametrized by the $N_H$ value. The 
corresponding $N_H$ value is then obtained by comparing the observed HR with the predicted ones. For mildly obscured sources, this is the same procedure 
followed by \citet{Marchesi:16} when there are fewer than 30 counts detected. For heavily obscured and CT sources, the observed X-ray spectrum can be more
complicated than the simple power law and photoelectric absorption assumed before \citep[e.g.,][]{Matt:00,Arevalo:14,Bauer:15}. Hence, we further consider the 
predicted $N_H$-HR relations using the physically-motived X-ray spectral libraries from \citet{Murphy:09}, the so-called MYTorus models, which were not 
considered by \citet{Marchesi:16}. The hence-derived $N_H$ values are presented in Table\,\ref{tab:Xray_prop} and Figure \,\ref{fig:hr}.

\begin{table}
\caption{Herschel-DOGs with an X-ray detection in the Chandra COSMOS Legacy survey.}
\begin{tabular}{|c|c|c|c|c|c|c|c|c|}
\hline
DOG ID &  X-ray ID & SED-fitting & ra    &    dec   &    redshift &  HR        &   N$_H$                    &    L$_{ir}$        \\
       &	& flag   &	   &          &             &            &   $\times$10$^{22}$ cm$^{-2}$    &    (L$_{\odot}$)    \\
 \hline \hline 
       9  &    lid 3606	&   host  &      149.932  &      1.626  &      1.42  &     0.75$^{+ 0.25}_{-0.07}$  &      58.4 --       927  &   7.2\,$\times$\,10$^{11}$  \\
      11  &   lid 2467	&    host  &      149.952  &      1.744  &      1.63  &    -0.36$^{+0.40}_{-0.29}$  &    0.01 --       12.7   &   2.0\,$\times$\,10$^{12}$  \\
      42  &   lid 4354 	&    host  &      149.478  &      2.133  &      1.58  &     0.12$^{+0.46}_{-0.32}$  &      4.46 --       36.3  &   9.8\,$\times$\,10$^{11}$  \\
      56  &   lid 2346	&     host  &      149.733  &      2.335  &      1.58  &    -0.90$^{+0.005}_{-0.11}$  &      0.0 --       0.0  &   1.1\,$\times$\,10$^{12}$  \\
      60  &   lid 3101	&    host  &      150.507  &      2.598  &      1.27  &     0.86$^{+0.14}_{-0.04}$  &      39.9 --       927  &   6.9\,$\times$\,10$^{11}$  \\
      73  &   lid 319	&     AGN  &      150.426  &      2.725  &      1.20  &   -0.04$^{+0.09}_{-0.09}$  &      5.39 --       8.69  &   5.0\,$\times$\,10$^{11}$  \\
      74  &   lid 3055  	&    AGN  &      150.021  &      2.775  &      2.09  &     0.62$^{+0.38}_{-0.08}$  &      30.0 --       927  &   7.3\,$\times$\,10$^{11}$  \\
      80  &   lid 3931	&    AGN  &      149.562  &      2.696  &      1.89  &     0.65$^{+0.26}_{-0.15}$  &      36.3 --       103  &   1.1\,$\times$\,10$^{12}$  \\
      81  &   lid 1806	&    host  &      149.682  &      2.652  &      2.34  &    -0.13$^{+0.23}_{-0.21}$  &      2.29 --       20.5  &   1.5\,$\times$\,10$^{12}$  \\
      95  &   lid 306	&    host  &      150.379  &      2.735  &      0.92  &    0.06$^{+0.09}_{-0.09}$  &      3.35 --       4.90  &   4.2\,$\times$\,10$^{11}$  \\
      96  &   cid 201	&    AGN  &      149.906  &      1.917  &      1.49  &    -0.25$^{+0.08}_{-0.09}$  &      1.07 --       3.35  &   6.7\,$\times$\,10$^{11}$  \\
      97  &   cid 817	&    host  &      150.063  &      1.945  &      2.15  &    -0.19$^{+0.33}_{-0.29}$  &    0.01 --       20.5  &   1.7\,$\times$\,10$^{12}$  \\
      98  &   cid 1467	&    host  &      149.837  &      1.972  &      1.02  &   -0.02$^{+0.38}_{-0.33}$  &     0.55 --       8.69  &   6.5\,$\times$\,10$^{11}$  \\
      99  &   lid 2663	&   --       &      149.779  &      1.586  &      1.24  &    0.08$^{+0.45}_{-0.31}$  &      2.08 --       16.9  &      -99  \\
     100  &   cid 1091	&    host  &      150.106  &      2.014  &      1.88  &    -0.16$^{+0.35}_{-0.24}$  &     0.34 --       16.9  &   2.8\,$\times$\,10$^{12}$  \\
     101  &   lid 1646	&    AGN  &      150.787  &      2.151  &      1.47  &    -0.17$^{+0.02}_{-0.02}$  &      3.35 --       3.68  &   3.7\,$\times$\,10$^{11}$  \\
     102  &   lid 1565	&    AGN  &      150.547  &      1.619  &      1.59  &    -0.30$^{+0.05}_{-0.06}$  &     0.88 --       2.52  &   6.5\,$\times$\,10$^{11}$  \\
     103  &   lid 3587 &    AGN  &      149.931  &      1.735  &      1.43  &     0.77$^{+0.15}_{-0.12}$  &      30.0 --       64.3  &   9.4\,$\times$\,10$^{11}$  \\
     104  &   cid 476 	&    host  &      150.475  &      2.094  &     0.56  &     0.21$^{+0.04}_{-0.04}$  &      3.35 --       3.35  &   6.0\,$\times$\,10$^{11}$  \\
     105  &   cid 593	&    host  &      150.472  &      2.324  &     0.89  &     0.64$^{+0.13}_{-0.12}$  &      11.6 --       16.9  &   3.3\,$\times$\,10$^{11}$  \\
     106  &   cid 92	&    host  &      150.288  &      2.382  &      1.58  &    -0.50$^{+0.25}_{-0.24}$  &    0.01 --       2.52  &   2.9\,$\times$\,10$^{12}$  \\
     107  &   cid 1917	&    host  &      149.998  &      2.578  &      2.42  &    0.03$^{+0.97}_{-0.07}$  &      24.8 --       927  &   4.3\,$\times$\,10$^{12}$  \\
 \hline 
 \end{tabular}
\label{tab:Xray_prop}

\medskip
NOTE: {\it SED-fitting flag} is based on far-IR SED decomposition (see \citealt{Riguccini:15} for details).
\end{table}

As can be seen, there are no major differences between the simple obscured power law and the MYTorus models for moderately obscured sources, $N_H$
$<$10$^{23}$~cm$^{-2}$, up to $z$$\sim$2, where most of our sources are located. However, as it is expected, MYTorus predict in general lower HR values 
(softer X-ray spectra) for CT sources. This implies that just using the hardness ratio it is hard to discriminate a heavily obscured $N_H$$>$5$
\times$10$^{23}$~cm$^{-2}$ from a CT, $N_H$$>$10$^{24}$~cm$^{-2}$, source. However, given the low number of counts detected for the X-ray sources 
in our sample, this procedure is the best we can do to attempt to identify CT AGN.

According to the X-ray classification based on the derived $N_H$ values, 6 out of the 22 X-ray detected DOGs (i.e. 27\%) are plausible CT AGNs, 
namely DOG\# 9, 60, 74, 80, 103 and 107. This is strictly an estimate, since as shown in Fig.\,\ref{fig:hr}, sources with a HR in the $\sim$0.6-0.8 range 
can either be moderately/heavily obscured at $N_H$$\sim$few$\times$10$^{23}$~cm$^{-2}$ or CT. Further, using this classification scheme, 15 are 
considered as moderately-obscured AGNs, while only one of the X-ray detected DOGs has a low HR (DOG 56) consistent with being unobscured. The 
fraction of CT AGNs that we find in our sample is in good agreement with previous reports. For example, \citet{Georgakakis:10} found that the X-ray spectral 
properties of a sample of ``low-redshift DOGs analogues'' are consistent with moderate levels of 
obscuration and found in their sample a similar fraction of moderately-obscured AGN than our work. \citet{Ricci:15} found that 27$\pm$4\% of their sample of 
834 AGNs selected from the 70-month Swift/BAT catalog in the local Universe corresponds to CT AGNs. This is somewhat larger than the value predicted 
by \citet{Aird:15} at low redshifts but still in good agreement with the report by \citet{Burlon:11} using a smaller sample of 200 AGNs. It would have
been reasonable to expect that DOGs should have a higher fraction of CT sources because by definition they have dustier host galaxies. However, these 
results, combined with the evidence presented on section  4.4 appear to indicate that there is no significant difference with the general AGN population. Thus, 
we can speculate that the obscuration, at least in the most extreme cases has to be nuclear and roughly independent of the properties of the host galaxy, as 
also concluded by \citet{Ricci:17} for a hard X-ray selected AGN sample and using statistical arguments by \citet{Buchner:17}.

\subsection{X-ray stacking}
\label{sec:stacking}

Previous studies have shown that stacking in the X-ray is a powerful technique that allows the detection of emission from objects lying below the formal detection limit 
for individual sources \citep[e.g.,][]{Brandt:01}. Chandra is particularly well suited for this thanks to its very low and stable background. We perform X-ray stacking for the DOGs in the area covered by the COSMOS Chandra Legacy data using the web-based CSTACK code\footnote{http://cstack.ucsd.edu/ or http://lambic.astrosen.unam.mx/cstack/} developed by Takamitsu Miyaji. Stacking was performed in two bands independently: soft (0.5--2 keV) and hard (2-- 5 keV). Chandra internal background being dominated by strong emission lines above 7\,keV, we limit the high energy band threshold to 5\,keV to limit the internal background. Prior to stacking, we removed all the sources that were individually detected by Chandra; this reduced our sample from 108 far-IR detected DOGs to 86. After removing DOGs that are too close to an X-ray source, the stacking with CSTACK was performed on 76 objects. The radius of the exclusion region varies with the off axis angle, corresponding to the 90\% encircled counts fraction radii, with a minimum of 1.0 arcsec and a maximum of 7.0 arcsec.
We obtain a mean count rate on the soft band of 8.56$\pm$1.89$\times$10$^{-6}$\,cts/s. In contrast, no significant detection is obtained in the hard band using a $\sim$ 3-$\sigma$ threshold. We provide for different stacking approaches an estimate for the flux only when the detection is above 3-$\sigma$ (see Table\,\ref{tab:subsamples_stacking}). 

In order to convert count rates into fluxes we use the Portable, Interactive Multi-Mission Simulator (PIMMS) tool for the Chandra Observatory\footnote{http://cxc.harvard.edu/toolkit/pimms.jsp}. Assuming the corresponding Chandra-Cycle 14\,/\,ACIS response functions, an intrinsic power-law spectrum with $\Gamma$ = 1.9, a Galactic absorption value of 2.6 $\times$ 10$^{20}$ cm$^{-2}$ \citep{Willingale:13} and a representative intrinsic absorption of 10$^{23}$ cm$^{-2}$ at $z$=2 (median redshift of our sample of X-ray undetected DOGs), we find a conversion factor from counts-per-second to flux of 5.2$\times$10$^{-12}$ erg cm$^{-2}$ s$^{-1}$/(cts\,s$^{-1}$) in the soft band. We find that the average observed X-ray flux for the X-ray undetected DOGs is S$_{0.5-2 keV}$=4.4$\times$10$^{-17}$\,erg\,cm$^{-2}$\,s$^{-1}$ in the soft band, a factor of 10 lower than the average value for the X-ray detected DOGs. The stacked signal in the hard band is below a 3-$\sigma$ detection (cf Table\,\ref{tab:subsamples_stacking}).

\section{Discussion}
\label{sec:discussion}

\subsection{X-ray Stacking of Specific Population Sub-samples}

In this section, we study the possible dependence of our X-ray stacking results on other parameters of the DOGs such as AGN activity, star formation, redshift and 24$\mu$m flux. Our results are described in this section and summarized in Table\,\ref{tab:subsamples_stacking}.

\begin{table}
\centering
\caption{Characteristics of the different subsamples of Herschel-DOGs from \citet{Riguccini:15} on which X-ray stacking has been performed with the C-STACK procedure }
\begin{tabular}{|c|c|c|c|c|c|c|c|c|}
\hline
Subsample &  Number of  &  Number of & median    &        Flux *        &   median   &           Flux   *       \\
 	           & sources used &  FIR-AGN DOGs &  count rate &      Soft band   & count rate   &       Hard band (2-8 keV)    \\
 	     &    for stacking &    used for stacking & Soft band   &     [erg/s/cm2]  &  2-5 keV   &         [erg/s/cm2]   \\
	     &                   &             &      (1e-06)	          &        (1e-17)   &   (1e-06)	          &        (1e-17)    \\
 \hline \hline
all undetected 	&     76         &     10   &    8.56 $\pm$ 1.89  &        4.43  &   6.01 $\pm$ 2.24         &        --       \\
SF undetected	&	59	   &	    0     &    7.39 $\pm$ 2.05  &        3.82    &    6.52  $\pm$ 2.55       &      --     \\
AGN undetected &   10        &     10    &   10.5 $\pm$ 6.45   &       --        &    4.24 $\pm$ 6.43      &       --            \\
R - K $>$ 2.79   &    69         &	  10    &    8.94 $\pm$ 2.02   &      4.62   &   7.24 $\pm$ 2.41         &   1.76 \\
 \hline 
SFR$>$200     &    55          &	   6      &    8.83 $\pm$ 2.18    &       4.57  &   5.03 $\pm$ 2.63        &         --        \\
SFR$>$300     &     38         &      5      &    10.1 $\pm$  2.86   &        5.65  &    4.72 $\pm$ 3.16       &         --  \\
SFR$>$400     &     23         &      4      &    5.63 $\pm$  3.40   &        --              &    1.94 $\pm$ 3.86       &     --       \\
 \hline 
 z$<$1.9          &     28         &      0       &   9.53 $\pm$ 3.01  &        4.94    &    7.18 $\pm$ 3.91        &        --         \\
 1.9$<$z$<$2.3 &   24         &      2       &   10.5 $\pm$ 3.42    &        5.43    &    8.70 $\pm$ 4.17         &      --       \\
 2.3$<$z$<$3  &     25         &      8       &    7.57 $\pm$ 3.41    &          --      &    7.22 $\pm$ 3.63         &       --       \\
 \hline 
 0.09$<$F24$<$0.36 &   30  &    1        &   7.14 $\pm$ 2.77    &       --    &    7.20 $\pm$ 3.56         &     --      \\
 0.36$<$F24$<$0.60 &   26  &    1        &   5.58 $\pm$ 3.27    &        --    &    4.08 $\pm$ 3.76      &       --       \\
 0.60$<$F24$<$4.74 &   21  &    8        &   15.6 $\pm$ 4.19    &      8.08    &    3.81 $\pm$ 4.45       &    --       \\
  \hline 
\end{tabular}
\\  * Fluxes are computed using PIMMS only when a detection $>$3-$\sigma$ significant is achieved. 
\label{tab:subsamples_stacking}
\end{table}

%
%

\subsubsection{Effect of AGN activity on X-ray stacking}

Following the far-IR classification of \citet{Riguccini:15}, we stacked the sub-sample of ``star-forming" DOGs (those dominated by a host SED component) and that of far-IR AGN DOGs. Out of the DOGs that are not detected in the X-rays, a total of 66 are classified as ``host" DOGs by \citet{Riguccini:15} and 12 are classified as AGNs; note that only 10 sources of the latter were used to perform the stacking after eliminating two due to proximity to an X-ray source. The stacked signal in the soft band for the ``host" DOGs is reported in Table\,\ref{tab:subsamples_stacking}. Most likely due to the low number of sources, no signal was 
detected either in the soft nor the hard band of the AGN far-IR DOGs. 

\citet{Fiore:08,Fiore:09} and \citet{Treister:09b} showed that imposing a color cut of R\,-\,K$\,>\,$2.79 on a DOG sample increases the probability of selecting AGN DOGs. It is worth noting that we find similar results with our X-ray stacking analysis. Indeed the only subsample where we obtained a stacked emission in the hard band (i.e. with detection $>$ 3\,$\sigma$) is for the DOG population with R-K$>$2.79, underlying a higher AGN activity in this subsample.
Based on this result, we estimate the hardness ratio of the stacked signal for the R-K cut sample and obtain a value of 0.02$\,\pm\,$0.29. Considering this result and the redshifts of the sources that went into the stacking, we can see from Figure\,\ref{fig:hr} that this contribution is likely associated with moderately-obscured AGN activity.

%
%

\subsubsection{Effect of redshift and star-formation activity on X-ray stacking}

In order to gauge the impact that redshift may have in our results, we divide our sample of 86 X-ray undetected DOGs roughly evenly into three redshift bins and perform X-ray stacking independently on these 3 sub-samples: 30 sources at z$<$1.9, 27 sources with 1.9$<$z$<$2.3 and 29 sources with 2.3$<$z$<$3. After excluding DOGs that are too close to an X-ray source the stacking was performed on 28 sources at z$<$1.9, 24 at 1.9$<$z$<$2.3 and 25 at 2.3$<$z$<$3, with corresponding median redshifts of $<$$z$$>$ $\sim$ 1.75, 2.0 and 2.7, respectively. Table \ref{tab:subsamples_stacking} displays our findings, where quoted fluxes are calculated assuming a conversion factor of  5.18\,$\times$\,10$^{-12}$~erg~cm$^{-2}$s$^{-1}$/(cts/s). The X-ray stacking procedure yielded significant detections (i.e., $>3$-$\sigma$) in the soft band for the two lower redshift bins, but not for the higher redshift bin probed in this study. We found no significant detections in the hard band for any of the redshift bins. Based on these results we do not find any evidence for a significant redshift evolution in the average soft X-ray flux of the sample.

We consider 3 bins of increasingly intense star formation activity $-$ SFR$>$200$M_\odot$yr$^{-1}$ (55 sources), SFR$>$300$M_\odot$yr$^{-1}$ (38 sources) and SFR$>$400$M_\odot$yr$^{-1}$ (23 sources) $-$ neither of which present a signal in the hard band. We merely find a detection in the soft band for the bins with sources displaying SFR$>$200$M_\odot$yr$^{-1}$ and SFR$>$300$M_\odot$yr$^{-1}$. Based on these results we are unable to probe for any trends with respect to star formation activity.

%
%

\subsubsection{Effect of 24$\mu$m flux on X-ray stacking}

To analyze the effect of the 24$\mu$m flux on the X-ray properties of the DOGs, we split our sample in three 24$\mu$m flux bins: 0.09$<$F24($\mu$Jy)$<$0.36, 0.36$<$F24($\mu$Jy)$<$0.60, 0.60$<$F24($\mu$Jy)$<$4.74 with mean redshifts of $<z>\sim$2.1, 2.1 and 2.2 respectively. We performed X-ray stacking on these three sub-samples independently and only find a $>3$-$\sigma$ detection for the brightest 24$\mu$m in the soft band. However, based on tentative detections ($<3$-$\sigma$) for the fainter 24$\mu$m bins, the soft band stacking results suggest a dependence on the 24$\mu$m flux, with higher X-ray fluxes associated to brighter 24$\mu$m sources. This is an expected trend, as earlier works \citep[e.g.,][]{Treister:06,Riguccini:15} have shown that the AGN fraction increases strongly with 8$\mu$m luminosity and hence with the 24$\mu$m flux as well.

 \subsection{AGN Fraction}

\begin{figure*}
 \centering
 \resizebox{0.7\hsize}{!}{\includegraphics{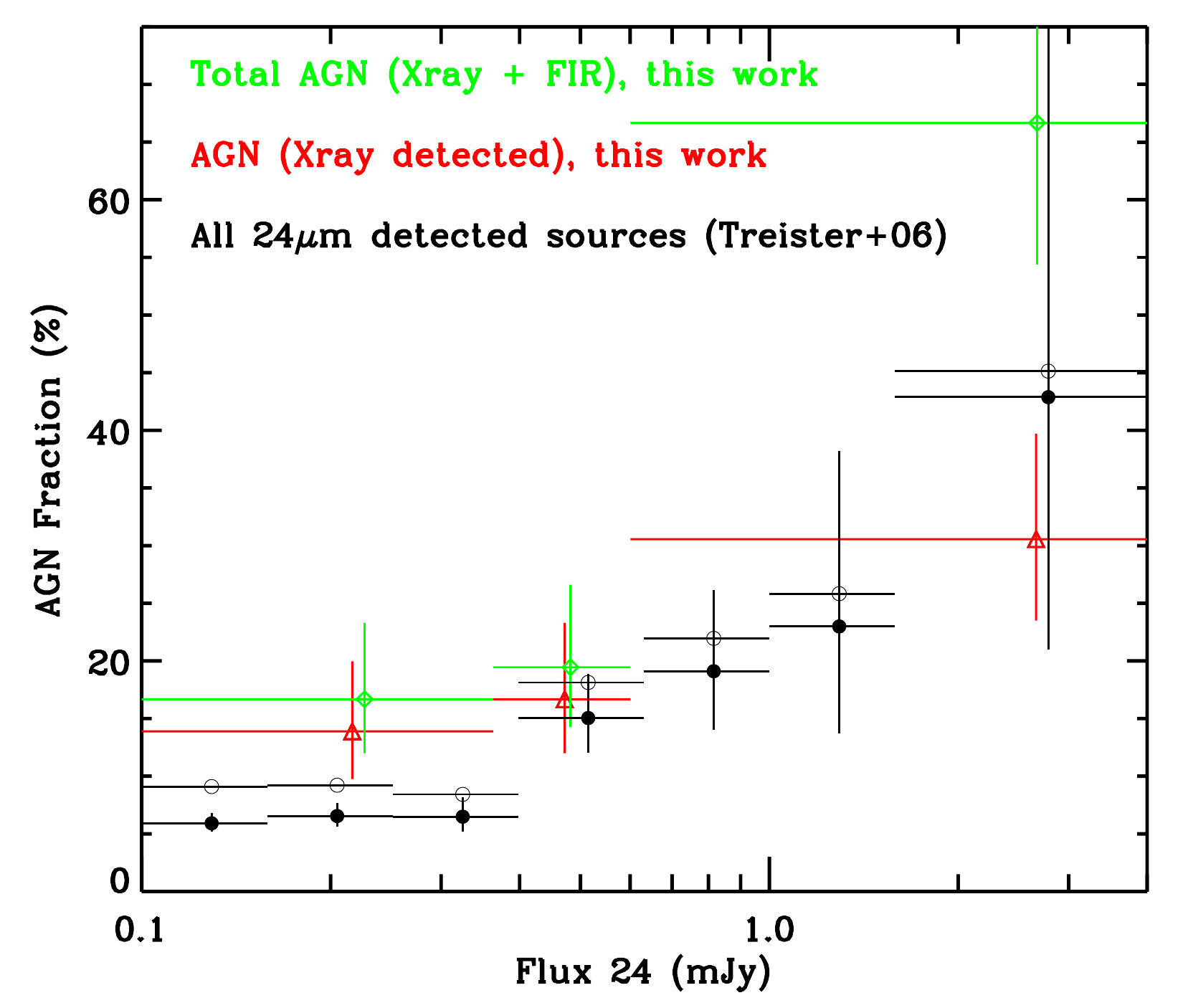}}
  \caption{Fraction of Herschel-DOGs classified as AGNs as a function of 24$\mu$m flux. The {\it green diamonds} show the combination of far-IR and X-ray selected 
  AGNs, while the {\it red triangles} show only the X-ray selected AGNs. The horizontal error bars show the size of the flux bins, while vertical error bars show the 
  1$\sigma$ Poissonian errors on the number of sources. As a comparison we show the fraction of sources classified as AGNs in the GOODS field 
  ({\it filled black circles}) and the fraction corrected by the AGNs expected to be missed by X-ray selection, as estimated using an AGN population synthesis model 
  ({\it open black circles}), as described by \citet{Treister:06}.}
 \label{fig:AGN_fraction_FL24}
\end{figure*}

We show in Fig.\ref{fig:AGN_fraction_FL24} the fraction of far-IR detected DOGs that are classified as AGNs as a function of the 24$\mu$m flux, considering only X-ray classification (red triangles) and the combined X-ray and far-IR SED-based classification (green diamonds). We can see that the AGN fraction increases rapidly towards higher 24$\mu$m fluxes, particularly when considering the combined far-IR and X-ray detected analysis. \citet{Brand:06} found that at the brightest 24$\mu$m fluxes, 74\%$\pm$21\% of their sample of LIRGs with 
z$>$0.6 have their mid-IR emission dominated by an AGN. 
Compared to the entire 24$\mu$m population in the GOODS field (see also, e.g., \citealt{Treister:06}),  we only see a significant difference between the DOGs and the wider 24$\mu$m population at the lowest 24$\mu$m fluxes. However, the X-ray-based AGN fraction shown in Fig.\ref{fig:AGN_fraction_FL24} is strictly based on considering merely those AGN DOGs with individual X-ray detections. In the light of our X-ray stacking results, we note that AGN activity is not limited to this sample of individually X-ray detected DOGs, but that a mix nature (AGN and star formation) exists within the population of the individually X-ray undetected DOGs. In an effort to constrain the contribution to the AGN fraction from the X-ray undetected population,  we compared our stacked point with results from X-ray normal galaxies from \citet{Lehmer:16} and with our X-ray detected DOGs and found a contribution of 20\% from AGN activity for the stacked sample. Considering that the X-ray undetected DOG sample in question is comprised of 76 DOGs, this translates into a potential increase of the AGN fraction by 15 significantly obscured DOGs.

Our stacking analysis showed that X-ray fluxes increase with 24$\mu$m flux. This is consistent with the observed trend within the X-ray detected DOG population, with 
an observed increase of the total AGN fraction in the brightest 24$\mu$m bins (see Fig.\ref{fig:AGN_fraction_FL24}), as previously reported by e.g., 
\citet{Dey:08} and \citet{Fiore:09}. Taking into account the potential non-negligible fraction of highly obscured AGN missed even with our (X-ray + far-IR) combined analysis 
but revealed within our undetected DOG sample, we expect that the AGN fraction traced by the DOG population may be even higher within the brightest 24$\mu$m 
bins. Combined with the observed difference in AGN fraction at the faintest 24$\mu$m bins between the DOG population and the GOODS 24$\mu$m population as 
a whole from \citet{Treister:06}, these results point to the DOG population as an effective means of selecting AGNs, particularly so in the case of high obscuration.

\subsection{Star Formation Rates}

\begin{figure}
 \centering
 \resizebox{1.\hsize}{!}{\includegraphics{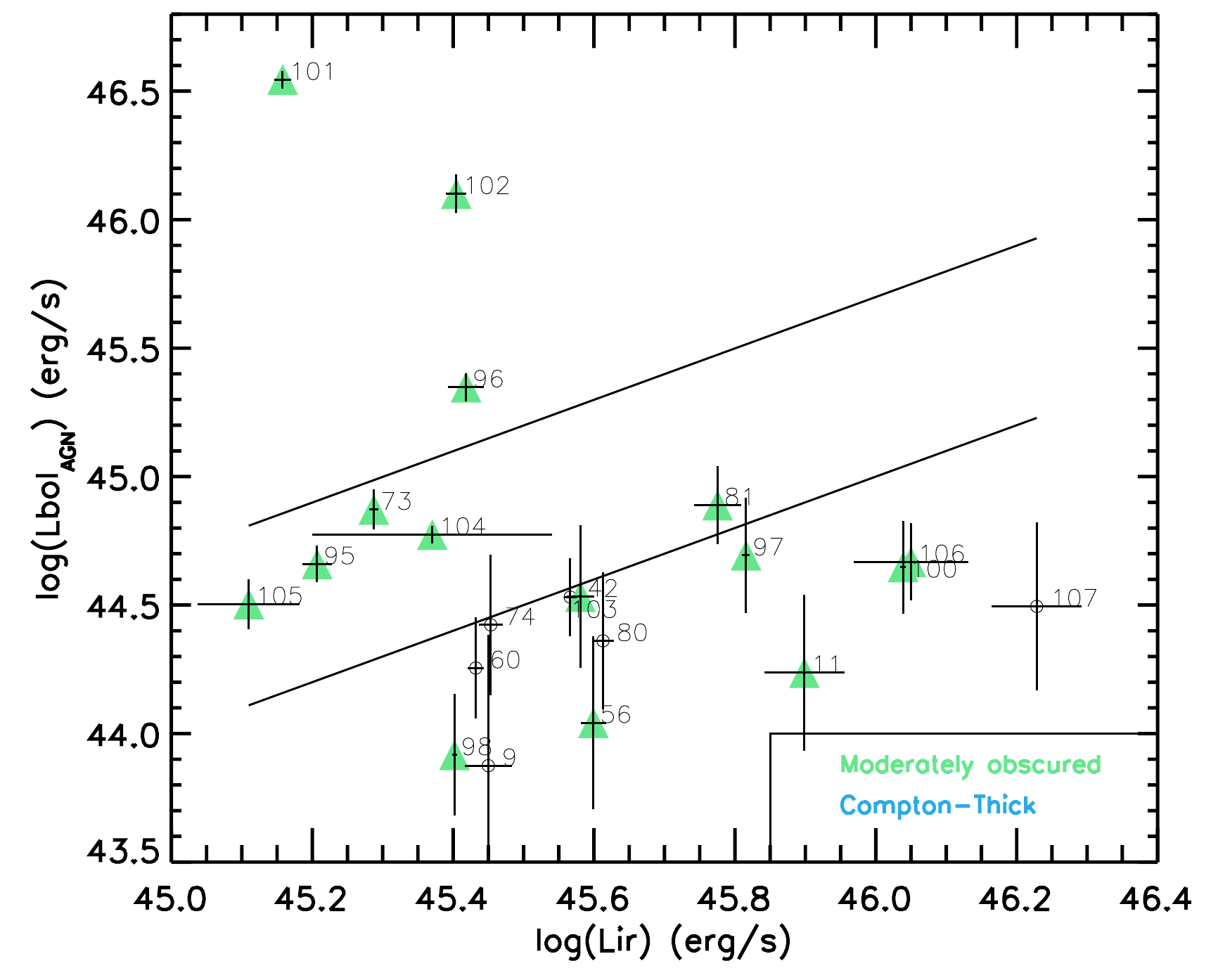}}
  \caption{AGN bolometric luminosity as a function of the infrared luminosity for the 21 X-ray detected DOGs (CT candidates: blue stars; moderately obscured 
  AGNs: green triangles) with a FIR SED-decomposition fit that led to the determination of the IR luminosity. Error bars in luminosities are obtained directly from the errors in the fluxes, accounting for the observed spread in bolometric correction in X-rays. The numbers next to the symbols correspond to the DOG ID\,\# from \citet{Riguccini:15}, also listed in Table\,\ref{tab:Xray_prop}. The lower and upper solid lines represent 
  10\% and 50\% of the infrared luminosity, respectively.}
 \label{fig:Lbol_lir}
\end{figure}

In order to identify a potential AGN contribution to the IR luminosity used to infer star formation rates, we first derive the AGN bolometric luminosity for the 22 X-ray detected DOGs (0.5-10\,keV), estimating it from the intrinsic (i.e., absorption-corrected) X-ray 
luminosity and assuming a fixed factor of 10 for the bolometric correction, as reported by  \citet[e.g.,][]{Rigby:09,Vasudevan:09}. While a luminosity dependence of the bolometric correction has been claimed in
the past \citep[e.g.,][]{Marconi:04}, more recent work \citep{Lusso:12} shows that in the luminosity range spanned by our sample the expected changes in
luminosity are relatively small, a factor of $\sim$2, and consistent with the observed dispersion, as can be seen in Figure 8 of \citet{Lusso:12}, thus justifying our
conservative choice of a constant bolometric correction. We then conclude that in most cases the AGN accounts for less than 50\% of the IR luminosity,
as can be seen in Fig.\,\ref{fig:Lbol_lir} for the 21 X-ray detected DOGs with a FIR-fit. This conclusion holds even considering a bolometric correction that is $\sim$2$\times$ higher for the most luminous 
sources. Hence, even in sources that contain an AGN, the nuclear emission does not make a significant contribution to the IR luminosity, which is most likely due to processes related to the star formation activity. In particular, we find that $\sim60\%$ of all moderately obscured AGNs and all CT candidates display a $\leq10$\% AGN contribution to the 
IR luminosity.

\begin{figure}
 \centering
 \resizebox{1.\hsize}{!}{\includegraphics{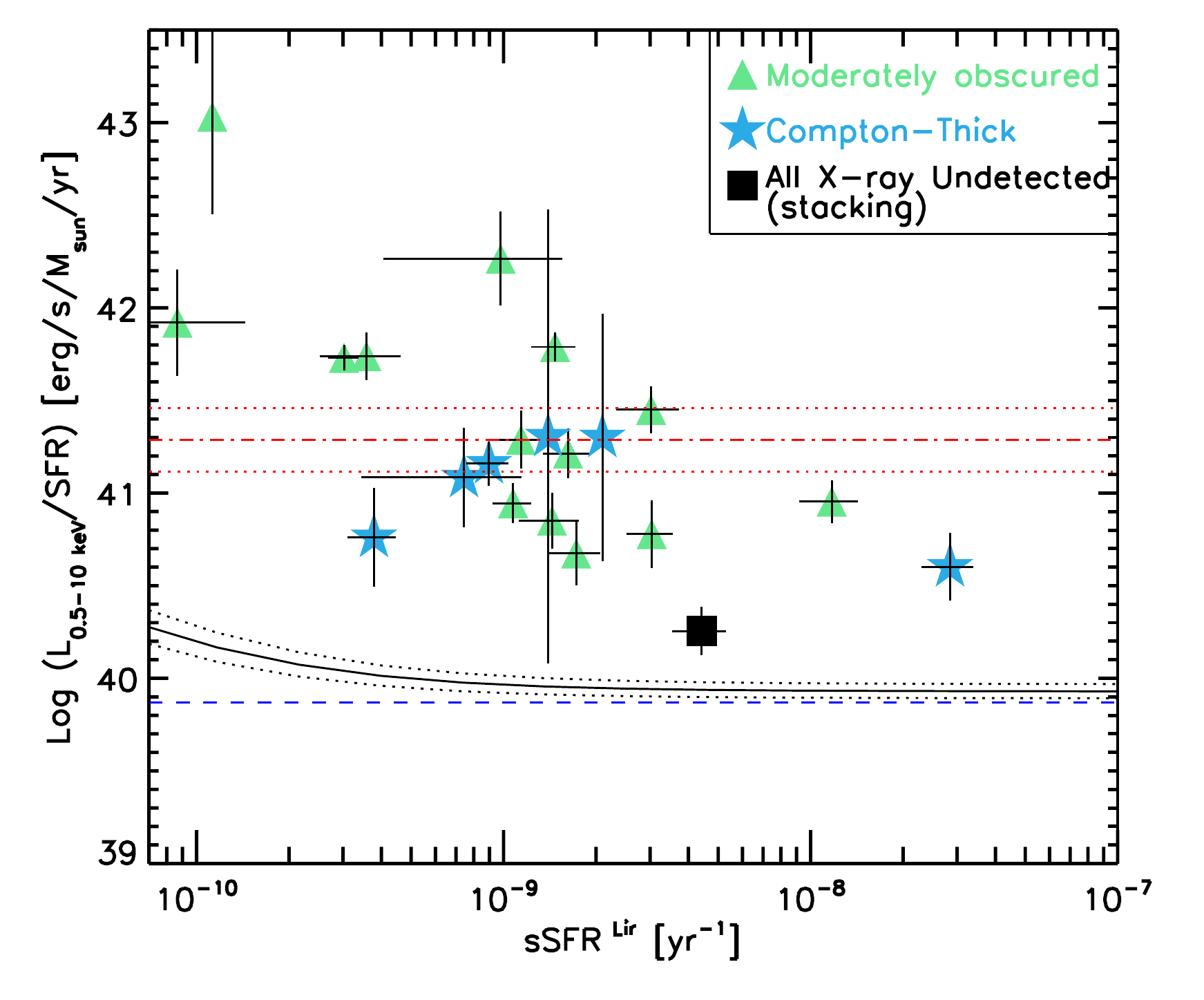}}
  \caption{Logarithm of the X-ray luminosity (0.5-10 keV) per unit SFR versus sSFR for the 21 X-ray detected DOGs (CT candidates: blue stars; moderately obscured 
  AGNs: green triangles) with FIR SED-decomposition fit. The black solid curve represent the best fit solution from the results of \citet{Lehmer:16} for 116 X-ray detected normal galaxies and the two dotted lines shows the error bars on this fit. The blue 
  dashed line shows the results from \citet{Mineo:14} for z$<$1.3 X-ray and radio  detected galaxies. 
  The red dotted-dashed line is the median value for the Xray detected sources and we are showing the median error bars with the two red dotted lines.
  The black rectangle shows the logarithm of the X-ray luminosity (0.5-10 keV) per unit SFR versus sSFR for the X-ray undetected sample (see sect.\,\ref{sec:stacking}).}
 \label{fig:Lx_SFR}
\end{figure}

Fig.\,\ref{fig:Lx_SFR} displays the X-ray luminosity per unit of star formation rate, in units of  erg s$^{-1}$/[$M_\odot$yr$^{-1}$], as a function of the specific SFR (sSFR) for the 
21 X-ray detected sources with FIR SED-decomposition fit. The stellar masses are taken from SED-fitting analyses using 30 bands in COSMOS from \citet{Ilbert:09}. The SFR is obtained using the \citet{Kennicutt:98} relation with IR luminosity. We use the IR luminosity obtained in this work with SED-fitting taking into account the contribution of an AGN component to the host galaxy. We are confident in our stellar masses and SFR estimates since they are in very good agreement with the recent work of \citet{Suh:17} on Type 2 AGN host galaxies in the Chandra COSMOS Legacy Survey. Our stellar masses (median log M$_{*}$ = 11.26$\pm$0.07) agree within the error bars with their value (median log M$_{*, Suh}$ = 11.00$\pm$0.20) and our median SFR (131$\pm$39\,M$_{\odot}$yr$^{-1}$) agrees with their median value as well (173$^{+45}_{-12}$\,M$_{\odot}$yr$^{-1}$).
The black dotted lines on Fig.\,\ref{fig:Lx_SFR} delimits the area populated by normal galaxies from \citet[][orange area]{Lehmer:16}. The fact that all of the sources in our sample are above the line found for normal (i.e., non-AGN) galaxies, as reported by \citet{Mineo:14},
in most cases by more than an order of magnitude indicates that the X-ray emission is most likely dominated by the AGN emission, even if the IR is not. 

We further include in Fig.\,\ref{fig:Lx_SFR} the results of our stacking analysis for all X-ray undetected DOGs. In order to do this, we used the PIMMS tool to convert the 
flux of the X-ray undetected DOGs stacked sample from Table\,\ref{tab:subsamples_stacking} into the extrapolated 5-10\,keV X-ray luminosity with a median redshift of z$=$2. The stacked sample has a median sSFR of 4.42$\pm$0.88$\times$10$^{-9}$\,[yr$^{-1}$].
We are quantifying the AGN contribution to the X-ray undetected DOG population using the stacked point in this figure. If the stacked sample was only star-formation activity, it would be located on the  black solid curve with log (Lx/SFR) $=$ 39.93\,[erg/s/M$_{\odot}$/yr] at the median sSFR of the Xray undetected sample . If it was only coming from AGN activity (i.e. 100\% contribution) it would be located on the red dashed-dotted line (i.e. log (Lx/SFR) =41.28\,[erg/s/M$_{\odot}$/yr]). We then estimate the AGN contribution of the stacked sample at $\sim$20\%. We note here that we found a similar AGN contribution using the R-K$>$2.79 X-ray undetected DOGs sample. 

\subsection{Specific star formation rate of the DOG population: where do the AGN-DOGs lie with respect to the main sequence?}

Recent studies have pointed to the existence of a so-called ``main sequence" (MS) of star forming galaxies up to $z\sim2.5$ where galaxies undergoing star 
formation activity typically lie within a well-defined region in the SFR-stellar mass diagram \citep[e.g.,][]{Tacconi:17, Genzel:15, Whitaker:12,Elbaz:11,Daddi:07}. \citet{Riguccini:15} showed, 
based on the MS definition of \citet{Elbaz:11}, that far-IR AGN DOGs mainly lie on or below the MS, while DOGs dominated by a host component lie on the MS and 
above it, within the starburst regime. These results underline the diversity found in the DOG population. Taking advantage of the {\it Herschel} data, we can 
derive, using SED-fitting at far-IR wavelengths, reliable IR luminosities and hence star formation rates, and potential AGN contributions in the IR for the DOGs 
in our sample.

\begin{figure*}
 \centering
  \resizebox{0.7\hsize}{!}{\includegraphics{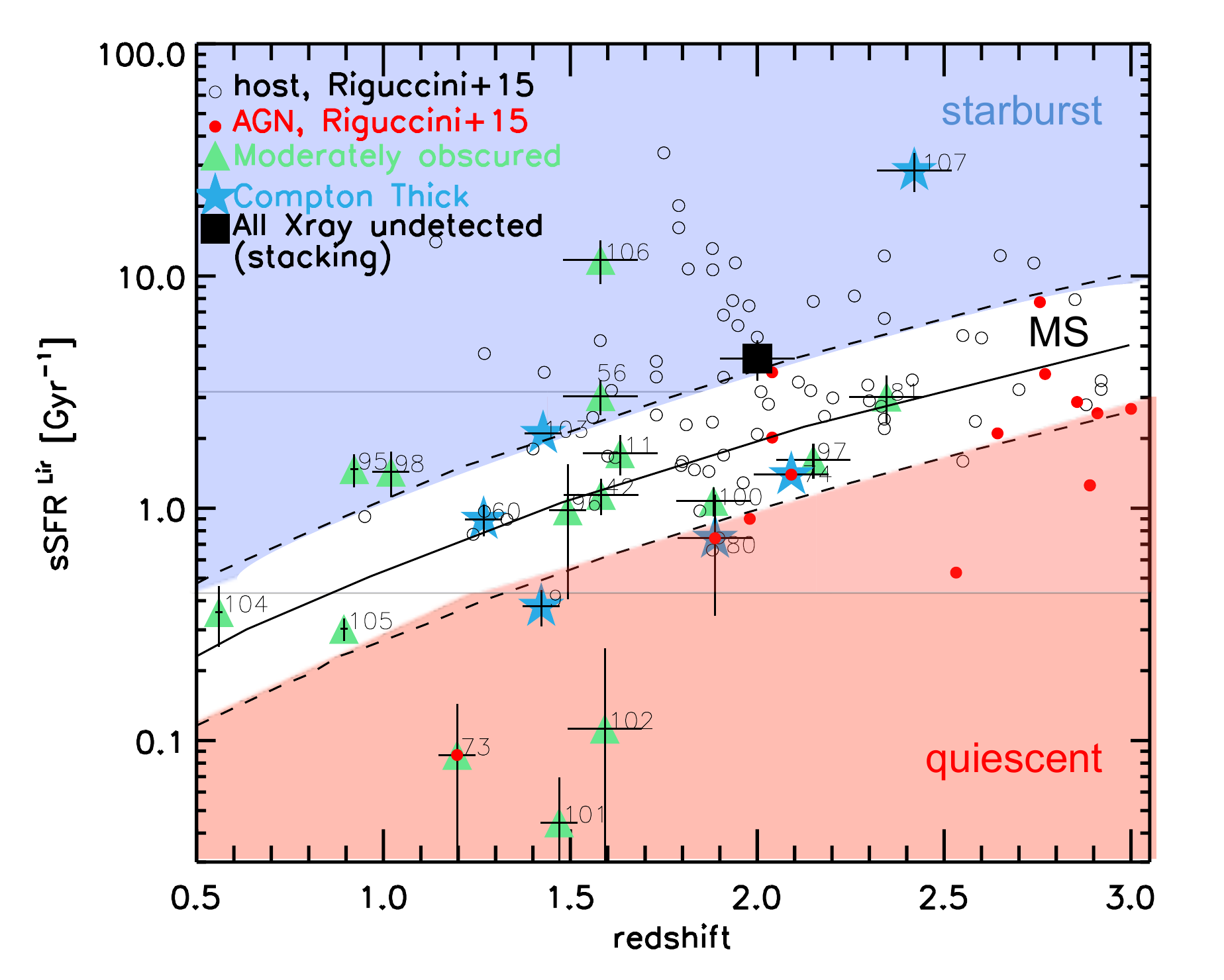}}
\caption{Redshift evolution of the specific SFR (sSFR = SFR/M$_*$) of detected in the infrared ({\it black} and {\it red circles}) and in X-rays (CT candidates: {\it blue stars}; heavily obscured AGNs: {\it green triangles}). The SFR rate is calculated from the infrared luminosity obtained from the SED-fitting procedure described by \citet{Riguccini:15}, i.e. removing the AGN contribution. The {\it solid line} represents the star forming main sequence from \citet{Elbaz:11}, while the {\it dashed lines} are a factor 2 above and below this fit. Most of the CT AGN ({\it blue stars}) are within the MS. The upper part of the plot corresponds to the starburst regime while the lower part corresponds to the quiescent phase. The black rectangle shows the location of the X-ray undetected DOG population (i.e. the stacking analysis, cf Section\,\ref{sec:stacking} for details) with respect to the MS.}
 \label{fig:SFR_xray}
\end{figure*}

Fig.\,\ref{fig:SFR_xray} shows the evolution of the specific SFR (sSFR) of DOGs with cosmic time. The AGN-DOGs (both X-ray and far-IR) appear to present a 
lower sSFR than SF-DOGs. We run a KS test to verify how distinct the distributions in sSFRs are and find a low probability of 1.6\,$\times$\,10$^{-4}$ that they arise from the same parent distribution. The majority of the AGNs-DOGs (both far-IR and X-ray) 
populate the area around and below the MS, with only two sources lying well above it: one CT candidate and a moderately obscured AGN. 

We do not see clear differences in behavior between the CT candidates and the moderately obscured AGNs in Fig.\,\ref{fig:SFR_xray}. According to the evolutionary scenario of \citet{treister:10}, the highly obscured CT AGNs correspond to the early, very dust-enshrouded, SMBH growth phase in a major galaxy merger; moderately obscured AGNs correspond in turn to a later stage in this evolutionary scenario, when the energetic feedback related to the SMBH accretion have already started heating up the dust and gas of the galaxy, shutting down star formation activity. In this picture, we expect the CT candidates to lie slightly above the moderately-obscured AGNs in the sSFR-redshift diagram, which does not appear to be the case. This surprising result that CT AGN are not preferentially found above the MS has been found as well in spectroscopically selected CT sample \citep[e.g.,][]{Georgantopoulos:13,Lanzuisi:15}. Hence, this further confirms the scenario presented on section 3.2, suggesting that at least for the most extreme sources, the obscuration has to be nuclear and thus not directly connected to the evolutionary stage of the host galaxy.

We note that the large uncertainties on the $N_H$ determinations make it hard to discriminate between CT and moderately obscured AGNs. This in turn also affects the derivation of the AGN contribution and have an impact in L$_{IR}$ and SFR estimates. This could potentially explain the lack of an observed difference of behavior between the two populations of X-ray detected AGNs. However, we have to consider the detectability of those sources. Assuming that the far-IR AGN have the same X-ray luminosity as the X-ray detected ones but they are lying at higher redshifts and hence have lower fluxes, we run C-stack on 19 random undetected DOGs to check if those far-IR AGN DOGs would have been detected by X-ray stacking, finding no significant detection.

%
%

\section{Conclusions}
\label{sec:ccl}

Searching for obscured Active Galactic Nuclei (AGN) is of main importance since AGN synthesis models for the Cosmic X-ray background (CXRB) predict a large number of obscured AGNs including CT AGNs. However, even the deepest Chandra and XMM-Newton surveys were able to detect only a few of them until nowadays. Our study combine exquisite new Chandra data with far-infrared Herschel data to catch obscured AGNs at z$\sim$2.
In this work we aimed to characterize the X-ray properties of the dust-obscured galaxy (DOG; F$_{24 \mu m}$/F$_{R}$$>$1000) population with far-IR detections by the Herschel Space Telescope. Our sample is composed of 108 DOGs in the COSMOS field and we relied on the Chandra COSMOS Legacy Survey X-ray and on the {\it Herschel} Observatory data to undertake our analysis. 

Out of 108 DOGs, 22 (i.e., 20\% of the sample) are individually detected in the X-ray soft and hard bands thanks to the increased coverage in area and sensitivity of the Chandra COSMOS Legacy Survey observations. Based on our estimates of the corresponding neutral hydrogen column density along the line of sight ($N_H$), we find that 6 of these X-ray detected DOGs (i.e., 27\%) are CT candidates, 15 (68\%) are moderately obscured AGNs and one is consistent with being unobscured. Our results are in excellent agreement with previous reports \citep[i.e.,][]{Ricci:15}, who found a fraction of CT AGNs of 27$\pm$4\%. This suggests that the fraction of CT sources is not different than that of the general AGN population and hence the obscuration, at least in the most extreme cases, appears to be independent of the amount of dust in the host likely and hence most likely nuclear.

We study the spectral energy distribution (SED) of the 22 X-ray detected DOGs, based on (rest-frame) optical-through-IR data, and find that only 7 are classified as AGN following the SED-fitting method described in \citet{Riguccini:15}. We note that out of the 19 far-IR AGN DOGs identified by \citet{Riguccini:15}, these 7 are the only ones with Chandra X-ray detection. This clearly shows how using far-IR to select obscured AGN is crucial to complement a X-ray analysis allow us to probe a wider range of AGNs.

Our main results are the following:

\begin{enumerate}

\item We find that the X-ray detected AGN DOGs and the far-IR AGN DOGs typically display similar near-IR and mid-to-far IR colors. Both populations are also typically found on the main-sequence of star-forming galaxies or below it. The main difference these populations appear to display is in their redshift distributions, with the far-IR AGN DOGs being typically found at larger distances. Together, these results suggest that the two populations share most of their physical properties and that the lack of detection in the X-ray band for the bulk of far-IR AGN DOGs is explained by the difference in redshift distributions. This strongly underline the critical need of multi-wavelength studies in order to obtain a more complete census of the obscured AGN population out to higher redshifts.

\item Based on earlier findings by \citet{Fiore:08,Fiore:09} and \citet{Treister:09b}, who showed that a color cut of R-K$>$2.79 on a DOG sample increases the probability of selecting AGNs, we stacked all individually-undetected DOGs above this color cut. This resulted in the strongest stacked signal from our sub-sample stacking, pointing to a higher AGN fraction, likely associated with moderately-obscured AGN activity.

\item We demonstrate that the combined population of X-ray detected and far-IR DOGs is effective at selecting AGNs, compared to the 24$\mu$m population as a whole \citep[as done within the GOODs field by e.g.,][]{Treister:06}. Moreover, X-ray stacking of individually-undetected DOGs points to a mix between AGN activity and star formation, where X-ray-undetected DOGs. We want to stress here how much our AGN far-IR {\it Herschel} SED-based classification is important. Indeed, if only considering X-ray detections, DOGs would have the same AGN fraction or even lower than a 24$\mu$m-selected population. This shows the critical need of deep far-IR surveys to probe AGN activity in star-forming galaxies samples.
\end{enumerate}

This work emphasizes the important role that the DOG population, in particular the combined X-ray and far-IR detected DOG population, plays in the effort to get a more complete census of the AGN population at high redshift, particularly for the highly obscured population. 



\acknowledgments

We would like to thank the anonymous referee for her/his very useful comments which significantly improved the paper. COSMOS is based on observations with the NASA/ESA Hubble Space Telescope, obtained at the Space Telescope Science Institute, which is operated by AURA, Inc., under NASA contract NAS 5-26555; also based on data collected at: the Subaru Telescope, which is operated by the National Astronomical Observatory of Japan; XMM-Newton, an ESA science mission with instruments and contributions directly funded by ESA Member States and NASA; the European Southern Observatory, Chile; Kitt Peak National Observatory, Cerro Tololo Inter-American Observatory, and the National Optical Astronomy Observatory, which are operated by the Association of Universities for Research in Astronomy (AURA), Inc., under cooperative agreement with the National Science Foundation; the National Radio Astronomy Observatory, which is a facility of the National Science Foundation operated under cooperative agreement by Associated Universities,Inc; and the Canada-France-Hawaii Telescope, operated by the National Research Council of Canada, the Centre National de la Recherche Scientifique de France, and the University of Hawaii. \\
PACS has been developed by a consortium of institutes led by MPE (Germany) and including UVIE (Austria); KU Leuven, CSL, IMEC (Belgium); CEA, LAM (France); MPIA (Germany); INAF-IFSI/OAA/OAP/OAT, LENS, SISSA (Italy); IAC (Spain). This development has been supported by the funding agencies BMVIT (Austria), ESA-PRODEX (Belgium), CEA/CNES (France), DLR (Germany), ASI/INAF (Italy), and CICYT/MCYT (Spain). \\
SPIRE has been developed by a consortium of institutes led by Cardiff University (UK) and including University of Lethbridge (Canada), NAOC (China), CEA, LAM (France), IFSI, University of Padua (Italy), IAC (Spain), Stockholm Observatory (Sweden), Imperial College London, RAL, UCL-MSSL, UKATC, University of Sussex (UK), Caltech, JPL, NHSC, University of Colorado (USA). This development has been supported by national funding agencies: CSA (Canada); NAOC (China); CEA, CNES, CNRS (France); ASI (Italy); MCINN (Spain); SNSB (Sweden); STFC, UKSA (UK) and NASA (USA).
SPIRE has been developed by a consortium of institutes led by Cardiff Univ. (UK) and including Univ. Lethbridge (Canada); NAOC (China); CEA, LAM (France); IFSI, Univ. Padua (Italy); IAC (Spain); Stockholm Observatory (Sweden); Imperial College London, RAL, UCL-MSSL, UKATC, Univ. Sussex (UK); Caltech, JPL, NHSC, Univ. Colorado (USA). This development has been supported by national funding agencies: CSA (Canada); NAOC (China); CEA, CNES, CNRS (France); ASI (Italy); MCINN (Spain); SNSB (Sweden); STFC, UKSA (UK); and NASA (USA). \\
ET acknowledges support from: CONICYT-Chile grants Basal-CATA PFB-06/2007 and AFB-170002, FONDECYT Regular 1160999 and 1190818, and  Anillo de ciencia y tecnolog\'ia ACT1720033. KMD and TSG thank the support of the Productivity in Research Grant of the Brazilian National Council for Scientific and Technological Development (CNPq). T. Miyaji and the development of CSTACK are supported by CONACyT  IB 252531 and UNAM-DGAPA PAPIIT IN104216 
E.L. is supported by a European Union COFUND/Durham Junior Research Fellowship (under EU grant agreement no. 609412) 
GL acknowledges support from the FP7 Career Integration Grant ``eEASy'' (CIG 321913) and grant ASI/INAF I/037/12/0-011/13. 
CR acknowledge financial support from FONDECYT 1141218, Basal-CATA PFB--06/2007 and the China-CONICYT fund.




\bibliographystyle{aasjournal} 

\bibliography{biblio_DOG_X}

\label{lastpage}



\end{document}